\documentclass[aps,nofootinbib]{revtex4}
\usepackage{amsmath}
\usepackage{graphicx}

\begin{document}

\title{Renormalisation-group analysis of repulsive three-body systems}
\author{Michael C. Birse}
\affiliation{Theoretical Physics Group, School of Physics and Astronomy,
The University of Manchester, Manchester, M13 9PL, UK}

\begin{abstract}
A coordinate space approach, based on that used by Efimov, is applied to 
three-body systems with contact interactions between pairs of particles.
In systems with nonzero orbital angular momentum or with asymmetric
spatial wave functions, the hyperradial equation contains a repulsive $1/r^2$ 
potential. The resulting wave functions are used in a renormalisation
group analysis. This confirms Griesshammer's power counting for short-range
three-body forces in these systems. The only exceptions are ones like the $^4S$
channel for three nucleons, where any derivatives needed in the interaction are 
found to be already counted by the scaling with the cut-off.

\end{abstract}
\maketitle

\vskip 10pt

Effective field theories (EFT's) are now being widely applied to few-nucleon 
systems, see Refs.~\cite{border,bvkrev}. The starting point is usually an
organisation of the terms in an effective potential according to naive 
dimensional analysis (NDA), as originally suggested by Weinberg \cite{wein}. 
This classifies terms according to number of powers of low-energy scales
they contain. In some cases, most notably $S$-wave nucleon-nucleon scattering, 
the leading-order (LO) terms turn out to be unnaturally large. This is a consequence
of low-energy bound or virtual states. It means that the LO terms need to be 
iterated to all orders in solving the Schr\"odinger or Lippmann--Schwinger 
equation \cite{bvk,vk,ksw}.

However it is now clear that NDA is not valid in 
all systems. There can be nonperturbative effects associated with strong 
long-range potentials that significantly change the power counting for 
short-range interactions. This was first noted in the context of attractive 
three-body systems (such as three bosons, or the $^2S$ channel for three 
nucleons) \cite{bhvk,bghr}, where the leading three-body forces must be promoted
to LO. More recently, failures of NDA have been observed in repulsive three-body 
systems \cite{grie}, and for nucleon-nucleon scattering in spin-triplet waves
\cite{ntvk,mcb}. In the first example, short-range three-body forces in most low 
partial waves are demoted to higher orders than naively expected; in the second, 
short-range forces are promoted to lower orders in $P$- and $D$-waves. In 
both cases, the terms in potentials scale with noninteger anomalous dimensions, 
and so the standard classification of terms as LO, next-to-leading order (NLO) 
etc.~is no longer convenient.

In this note I examine repulsive three-body systems using the renormalisation
group (RG) approach developed in Refs.~\cite{bmr,bb1,bb2}. This provides an
independent confirmation of results of Griesshammer for the power counting in 
these systems \cite{grie}. That work solved the Skorniakov--Ter-Martirosian equation 
\cite{stm} in momentum space, whereas here I work in coordinate-space following 
the approach developed by Efimov for attractive three-body systems \cite{efimov}
and recently extended by Gasaneo and Macek to cases with nonzero angular momentum 
\cite{gm}. I also comment on differences between the counting for derivative 
interactions in systems with strong long-range forces compared with the pure 
short-range case. This corrects the counting in Ref.~\cite{grie} for the leading 
three-body force in the $^4S$ channel for three nucleons.

If particles interact only through zero-range forces, then their wave functions 
satisfy the free Schr\"odinger equation, except where two of them coincide.
The two-body forces can then be represented by boundary conditions at these
points. These boundary conditions form the basis for Efimov's approach
\cite{efimov} as well more recent work in Refs.~\cite{fj,gom,gm}. In particular
Gasaneo and Macek have used this method to find solutions for systems with 
symmetric spatial wave functions. Here I generalise their results to cover
asymmetric cases, such as the spin-quartet channels for three nucleons. It is 
convenient to work in hyperspherical coordinates since the boundary conditions are
separable in the limit of infinite two-body scattering length. The resulting
hyperradial equation then has the form of a free radial Schr\"odinger equation 
with a centrifugal-like $1/r^2$ term whose strength is given by the hyperangular 
eigenvalue. This potential determines the form of three-body wave functions at 
small hyperradii, and hence it controls RG flow of the short-range three-body 
forces \cite{bb1,bb2}.
 
Sets of relative coordinates for three particles with equal masses are
\begin{equation}
{\bf x}_i={\bf r}_k-{\bf r}_j,\qquad {\bf y}_i={\bf r}_i-\frac{1}{2}\,
\left({\bf r}_j+{\bf r}_k\right),
\end{equation}
where $i,j,k$ are a cyclic permutation of 1,2,3, and I have used the 
traditional ``odd-man-out" notation to label the sets. 
Hyperspherical coordinates can then by defined in terms of these as
\begin{equation}
r=\sqrt{x_i^2+\frac{4}{3}\,y_i^2},\qquad \alpha_i=\arctan\left(
\frac{\sqrt{3}}{2}\,\frac{x_i}{y_i}\right).
\end{equation}

The three-body wave function can be decomposed into Faddeev components
\cite{fadd} as
\begin{equation}
\Psi(r,\Omega)=\sum_{i=1}^3 \frac{2}{r^2\sin(2\alpha_i)}\,
\phi_i(r,\alpha_i,\widehat{\bf x}_i,\widehat{\bf y}_i),
\label{eq:fc}
\end{equation}
where a factor of $1/(x_iy_i)$ has been taken out to simplify the radial parts
of the Hamiltonian. Away from the configurations where two particles coincide,
each of these components satisfies a free Schr\"odinger equation. In 
hyperspherical coordinates, this has the form
\begin{equation}
-\,\frac{1}{M}\left[\frac{\partial^2}{\partial r^2}
+\frac{1}{2}\,\frac{\partial}{\partial r}
+\frac{1}{r^2}\,\frac{\partial^2}{\partial \alpha_i^2}\right]\phi_i
+\frac{1}{Mr^2}\left[\frac{{\bf L}_i^2}{\cos^2\alpha_i}
+\frac{{\bf L}_{jk}^2}{\sin^2\alpha_i}\right]\phi_i=E\phi_i.
\end{equation}
Here ${\bf L}_{jk}$ denotes the relative angular momentum of the pair $jk$, and 
${\bf L}_i$ the angular momentum of particle $i$ relative to that pair.
If the problem is separable, we can write $\phi_i$ in the form
\begin{equation}
\phi_i(r,\alpha_i,\widehat{\bf x}_i,\widehat{\bf y}_i)=F_i(r)u_i(\alpha_i)
Y_{l_i^\prime m_i^\prime}(\widehat{\bf x}_i)Y_{l_i m_i}(\widehat{\bf y}_i),
\end{equation}
where $F_i(r)$ and $u_i(\alpha_i)$ satisfy the ordinary differential equations
\begin{eqnarray}
-\,\frac{1}{M}\left[\frac{{\rm d}^2}{{\rm d}r^2}
+\frac{1}{r}\,\frac{{\rm d}}{{\rm d}r}
-\frac{\nu^2}{r^2}\right]F_i&=&p^2F_i,\cr
\noalign{\vspace{5pt}}
-\,\frac{{\rm d}^2 u_i}{{\rm d}\alpha_i^2}+\left[\frac{l_i(l_i+1)}{\cos^2\alpha_i}
+\frac{l_i^\prime(l_i^\prime+1)}{\sin^2\alpha_i}\right]u_i&=&\nu^2u_i.
\end{eqnarray}
This hyperradial equation looks just like a free radial Schr\"odinger equation
in two dimensions, with the hyperangular eigenvalue $\nu^2$ determining the strength
of the centrifugal-like $1/r^2$ term. In the cases of interest, where pairs of 
particles interact only in $S$-waves, we can simplify the hyperangular equations by 
setting $l_i^\prime=0$ and $l_1=l_2=l_3\equiv l$. The component $\phi_i$ is then 
independent of $\widehat{\bf x}_i$.

From the definition of the reduced Faddeev components $\phi_i$ in Eq.~(\ref{eq:fc}), 
they must vanish at $x_i=0$. In hyperspherical coordinates these boundary conditions
are
\begin{equation}
\phi_i\!\left(r,\frac{\pi}{2},\widehat{\bf y}_i\right)=0.
\end{equation}
In the limit of infinite two-body scattering length, the logarithmic derivative
of the reduced wave function must vanish whenever two particles coincide and so can 
interact via the two-body force\cite{efimov,fj,gom,gm}:
\begin{equation}
\left.\frac{1}{x_i\Psi}\,\frac{\partial}{\partial x_i}\left(x_i\Psi\right)
\right|_{x_i=0}=0.
\end{equation}
The points where $x_i=0$ correspond to ${\bf x}_j={\bf y}_i$, 
${\bf y}_j=-\frac{1}{2}{\bf y}_i$ and ${\bf x}_k=-{\bf y}_i$,
${\bf y}_k=-\frac{1}{2}{\bf y}_i$ in the other relative coordinate systems.
In terms of the hyperangular coordinates these are $\alpha_j=\alpha_k=\frac{\pi}{3}$.
The resulting boundary conditions on the Faddeev components are
\begin{equation}
\left.\frac{\partial\phi_i}{\partial\alpha_i}\right|_{\alpha_i=0}
+\left.\frac{2}{\sin(2\alpha_j)}\,\phi_j\right|_{\alpha_j=\pi/3,\,
\widehat{\bf y}_j=-\widehat{\bf y}_i}
+\left.\frac{2}{\sin(2\alpha_k)}\,\phi_k\right|_{\alpha_k=\pi/3,\,
\widehat{\bf y}_k=-\widehat{\bf y}_i}=0.
\end{equation}
These are separable and lead to the hyperangular conditions
\begin{equation}
\left.\frac{{\rm d}u_i}{{\rm d}\alpha_i}\right|_{\alpha_i=0}
+\frac{4}{\sqrt 3}\,(-1)^l\left[u_j\!\left(\frac{\pi}{3}\right)
+u_k\!\left(\frac{\pi}{3}\right)\right]=0.
\end{equation}

The symmetries of the spatial wave function can be used to simplify these
conditions further. There are two cases of physical interest. The first is 
a spatial wave function that is symmetric under exchange of any pair of particles.
In this case the Faddeev components are equal:
\begin{equation}
\phi_1=\phi_2=\phi_3\equiv\phi.
\end{equation}
This describes three identical bosons, or three fermions with different 
quantum numbers (spin, isospin, etc.) whose intrinsic state is completely
antisymmetric. Most importantly for nuclear physics this corresponds to three
nucleons with total spin $\frac{1}{2}$. The second is where the spatial wave 
function is antisymmetric under exchange of one pair of particle, but symmetric 
under exchange of either of the others. The components are then related by
\begin{equation}
\phi_1=0,\qquad \phi_2=-\phi_3\equiv \phi.
\end{equation}
This describes three nucleons with total spin $\frac{3}{2}$. Generically we may 
write the boundary conditions in the form
\begin{equation}
\left.\frac{{\rm d}u}{{\rm d}\alpha}\right|_{\alpha=0}+\lambda (-1)^l\,
\frac{8}{\sqrt{3}}\,u\!\left(\frac{\pi}{3}\right)=0,
\label{eq:habc2}
\end{equation}
where $\lambda=+1$ for completely symmetric spatial wave functions 
and $\lambda=-\frac{1}{2}$ for cases with one antisymmetric pair.

As noted by Gasaneo and Macek \cite{gm}, the hyperangular equation for $l'=0$ 
can be solved in terms of hypergeometric functions. After defining the new
variable $z=\cos^2\alpha$ and writing $u(z)=z^{(l+1)/2}v(v)$, the equation
takes the form of the hypergeometric equation \cite{as},
\begin{equation}
z(1-z)\frac{{\rm d}^2v}{{\rm d} z^2}+[c-(a+b+1)z]\frac{{\rm d}v}{{\rm d} z}
-abv=0,
\end{equation}
with 
\begin{equation}
a=\frac{l+1-\nu}{2},\qquad b=\frac{l+1+\nu}{2},\qquad c=l+\frac{3}{2}.
\end{equation}
The hyperangular eigenfunctions are thus\footnote{Note that I have to chosen to 
write these in the form that will make most direct contact with the results of 
Ref.~\cite{grie}, rather than the equivalent form given in Ref.~\cite{gm}.
Also, the factor of 2 in Eq.~(13) of Ref.~\cite{gm} is incorrect and should be
omitted.}
\begin{equation}
u(\alpha)=(\cos\!\alpha)^{l+1}\,{}_2F_1\!\left(\frac{l+1-\nu}{2},\frac{l+1+\nu}{2},
l+\frac{3}{2};\cos^2\!\alpha\right).
\end{equation}
The corresponding radial solutions are just Bessel functions of order $\nu$
and so  we get
\begin{equation}
\phi(r,\alpha,\widehat{\bf y}_i)=N J_\nu(pr)\, (\cos\!\alpha)^{l+1}\,
{}_2F_1\!\left(\frac{l+1-\nu}{2},\frac{l+1+\nu}{2},l+\frac{3}{2};
\cos^2\!\alpha\right)Y_{lm}(\widehat{\bf y}_i).
\end{equation}

These vanish at $\alpha=\frac{1}{2}$, as required by the first boundary 
condition. The other condition, arising from the contact interactions at $x_i=0$,
then provides an equation for the eigenvalues $\nu^2$. The hypergeometric functions 
have the properties \cite{as},
\begin{equation}
\frac{\partial}{\partial z}\,{}_2F_1(a,b,c;z)=\frac{ab}{c}\,{}_2F_1(a+1,b+1,c+1;z),
\end{equation}
and for $z$ equal to or close to one
\begin{eqnarray}
{}_2F_1(a,b,c;1)&=&\frac{\Gamma(c)\Gamma(c-a-b)}{\Gamma(c-a)\Gamma(c-b)}\qquad
\mbox{if}\ c-a-b>0,\cr
\noalign{\vspace{5pt}}
{}_2F_1(a,b,c;z)&\sim&\frac{\Gamma(c)\Gamma(a+b-c)}{\Gamma(a)\Gamma(b)}\qquad
\mbox{if}\ c-a-b<0,
\end{eqnarray}
Using these, Eq.~(\ref{eq:habc2}) can be expressed in the form
\begin{equation}
1=\lambda\left(-\,\frac{1}{2}\right)^l\frac{2}{\sqrt{3\pi}}\,
\frac{\Gamma\left(\frac{l+1-\nu}{2}\right)\Gamma\left(\frac{l+1+\nu}{2}\right)}
{\Gamma\left(l+\frac{3}{2}\right)}\,{}_2F_1\!\left(\frac{l+1-\nu}{2},
\frac{l+1+\nu}{2},l+\frac{3}{2};\frac{1}{4}\right),
\end{equation}
which matches Eq.~(2.18) of Ref.~\cite{grie}, with the substitution of $s$ by $\nu$.
It also agrees with the results of Ref.~\cite{gm} if $\lambda=1$. 
For symmetric systems ($\lambda=1$) with $l=0$ this is the equation first derived by 
Danilov \cite{danil}, which has an imaginary solution for $\nu$. The corresponding 
hyperradial wave functions show oscillatory behaviour at small distances, and this 
is responsible for Thomas \cite{thomas} and Efimov \cite{efimov} effects. 
For $l\ge 1$ or systems with one antisymmetric pair, the roots of the equation are 
real and can be found in Table 2 of Ref.~\cite{grie}. In these cases the $1/r^2$
potential is repulsive and so there is no Efimov effect.

Having constructed the wave functions for the long-range forces in these systems, 
we can now use the methods of Ref.~\cite{bb1} to find the RG eigenvalues, which give 
the power counting for terms in the short-range potential. In fact for real values 
of $\nu$, we can immediately use the results in Eq.~(54) of that paper if we
multiply the hyperradial solutions by $\sqrt{\pi/(2pr)}$ to get functions that 
satisfy a three-dimensional radial Schr\"odinger equation. A term in the rescaled 
potential proportional to $p^{2n}$ ($n$ powers of the energy) varies with the 
cut-off $\Lambda$ as $\Lambda^{2(n+\nu)}p^{2n}$ and so its RG eigenvalue is 
\begin{equation}
\rho=2(n+\nu).
\end{equation}
If we assign $\Lambda$-independent terms to LO in our expansion of the EFT, then 
$\rho$ also labels the order of a term. The leading term in each channel is thus 
of order $2\nu$, in agreement with the results in Table 3 of Ref.~\cite{grie}, except 
for the $^4S$ and Wigner-antisymmetric $^2S$ channels, where Griesshammer adds two
extra powers of low-energy scales.

The motivation for adding these two powers is the antisymmetry of the wave function
in these channels which prevents all three particles from coinciding. As a result 
a pure $\delta$-function interaction has no effect on them; at one with at
least two derivatives would be needed. For any finite cut-off, however, a contact
interaction becomes nonlocal and so can contribute. In Ref.~\cite{grie}, this
happens  implicitly through the momentum cut-off. In contrast, Ref.~\cite{bb1} does 
it explicitly by using a $\delta$-shell form for the short-distance interactions.
This was done to ensure that the interaction has an effect even though the wave 
functions vanish as $r\rightarrow 0$ as a result of the $1/r^2$ potential.
For example, the same RG analysis can be applied to two-body scattering with nonzero 
angular momentum $L$ by setting $\nu=L+\frac{1}{2}$ \cite{bb1}. It shows that
the leading short-distance interaction in this partial wave is of order $2L$, as
expected from the fact that $2L$ derivatives of a $\delta$-function are needed to
form a contact interaction that acts in this wave. Note that these derivatives are 
already counted by the RG eigenvalue, $\rho=2L+1$. 

The wave functions in two-body channels with nonzero angular momentum, or in 
three-body channels with $\nu>-\frac{1}{2}$, satisfy a radial equation of the form
\begin{equation}
\frac{1}{r}\,\frac{{\rm d}^2}{{\rm d}r^2}(r\psi)=-p^2\psi
+\frac{\nu^2-\frac{1}{4}}{r^2}\,\psi.
\end{equation}
Acting on one of the wave functions where the long-range $1/r^2$ interaction has been 
iterated to all orders, an interaction with two derivatives thus gives rise to two
contributions. One is just proportional to two powers of the low-energy scale $p$, 
and so is two orders higher in the power counting. However the other, proportional
to $(1/r^2)\psi$, is of the same order as the term without those derivatives, since at 
small distances $1/r$ is not a low-energy scale. This second piece is absent if $L=0$ 
(or equivalently $\nu=\frac{1}{2}$). In that more familiar case, additional 
derivatives do indeed increase the order of the interaction. 

The bottom line is that any derivatives needed to construct 
appropriate interactions for the repulsive $1/r^2$ potentials are already counted in
the RG eigenvalue (or by the superficial degree of divergence in Ref.~\cite{grie}), 
without any need to add additional powers. But, apart from this rather minor 
amendment, the present analysis confirms Griesshammer's results for the power 
counting in repulsive three-body systems Ref.~\cite{grie}. The leading term in
each channel has RG eigenvalue $2\nu$, where $\nu^2$ in the hyperangular 
eigenvalue. These eigenvalues are not integers and so the usual classification
of terms as NLO etc.~in the EFT becomes inconvenient. 
In most cases this counting demotes short-distance three-body interactions to
higher orders than predicted by NDA, although in some channels there is a small
degree of promotion.

\section*{Acknowledgments}

I am grateful to the organisers of the ECT* Workshop on ``Nuclear forces and
QCD: never the twain shall meet?" where this work was started. I also thank H. 
Griesshammer for useful discusions and correspondence.

\end{document}